\documentstyle[12pt]{article}
\newlength{\dinwidth}
\newlength{\dinmargin}
\newcommand{\ba}{\begin{array}}
\newcommand{\ea}{\end{array}}
\newcommand{\be}{\begin{equation}}
\newcommand{\ee}{\end{equation}}
\newcommand{\bea}{\begin{eqnarray}}
\newcommand{\eea}{\end{eqnarray}}

\def\parallel{| \hskip-0.03cm |}

\newcommand{\tr}{\mbox{Tr}}

\def\bee{\begin{eqnarray}}
\def\eee{\end{eqnarray}}
\def\be{\begin{equation}}
\def\ee{\end{equation}}

\begin{document}
\thispagestyle{empty}
\addtocounter{page}{-1}
\begin{flushright}
SLAC-PUB-7396\\
SNUTP 96-130\\
SUITP 96-61\\
{\tt hep-th/9701139}\\
\end{flushright}
\vspace*{1.3cm}
\centerline{\Large \bf M(atrix) Theory on an Orbifold and Twisted Membrane
\footnote{
Work supported in part by the Department of Energy Contract
DE-AC03-76SF00515, NSF Grant PHY-9219345, U.S.NSF-KOSEF Bilateral Grant,
KOSEF Purpose-Oriented Research Grant 94-1400-04-01-3 and SRC-Program,
Ministry of Eduction Grant BSRI 97-2410
and the Seoam Foundation Fellowship.}}
\vspace*{1.7cm} \centerline{\large\bf Nakwoo Kim${}^a$ and
Soo-Jong Rey${}^{a,b,c}$}
\vspace*{0.4cm}

\centerline{\large\it Physics Department, Seoul National University,
Seoul 151-742 KOREA${}^a$}
\centerline{\large\it Department of Physics, Stanford University,
Stanford CA 94305 USA${}^b$}
\centerline{\large \it Stanford Linear Accelerator Center, Stanford
University, Stanford CA 94309 USA${}^c$}
\vspace*{2.1cm}

\centerline{\Large\bf Abstract}
M(atrix) theory on an orbifold and classical two-branes therein are
studied with particular emphasis to heterotic M(atrix) theory on
${\bf S}_1/ {\bf Z}_2$ relevant to strongly coupled heterotic and
dual Type IA string theories.
By analyzing orbifold condition on Chan-Paton factors, we show that
three choice of gauge group are possible for heterotic M(atrix) theory:
$SO(2N)$, $SO(2N+1)$ or $USp(2N)$. By examining area-preserving
diffeomorphism that underlies the M(atrix) theory, we find that
each choices of gauge group restricts possible topologies of
two-branes. The result suggests that only the choice of $SO(2N)$ or
$SO(2N+1)$ groups
allows open two-branes, hence, relevant to heterotic M(atrix) theory.
We show that requirement of both {\sl local } vacuum energy cancellation
and of worldsheet anomaly cancellation of resulting
heterotic string identifies supersymmetric
twisted sector spectra with sixteen fundamental representation spinors
from each of the two fixed points.
Twisted open and closed two-brane configurations
are obtained in the large N limit.
\vspace*{1.1cm}


\newpage

\section{Introduction}
\setcounter{equation}{0}

Witten~\cite{witten} has made an important observation that strong coupling
limit of perturbative string theories are unified into M-theory with an
eleven-dimensional Lorentz invariance. The massless excitations are
described by eleven-dimensional supergravity supplemented by additional
interactions necessary for anomaly cancellation.
BPS spectra and duality symmetries of various string theories have been
identified by studying macroscopic BPS two-branes and five-branes wrapped on
compactified space~\cite{schwarz}.
Despite such progress, a complete understanding of the M-theory has
remained elusive.

Recently, an interesting proposal~\cite{banksetal} have been put forward for
a non-perturbative definition of the M-theory.
This, so-called M(atrix) theory provides a partonic description of M-theory,
in  which the partons are type IIA 0-branes and open strings connecting them.
In the infinite momentum limit, the total energy of
0-brane partons grow arbitrarily large, hence, its number and transverse
density increases indefinitely. Correspondingly, the open strings connecting
them become arbitrarily short and leaves only massless gauge fields.
Thus, M(atrix) theory is a gauge theory of area-preserving diffeomorphism
among the 0-brane partons. As such, two-branes arise quite naturally as
collective excitations of 0-branes in a form of an incompressible fluid.

Horava and Witten\cite{horavawitten} have shown that M-theory compactified
on orbifold ${\bf S}_1/{\bf Z}_2$ along 11-th direction is a strongly coupled
$E_8 \times E_8$ heterotic string theory. At the two fixed points are located
the nine-branes or "ends of the world". Twisted sector of the M-theory
consists of twisted two-branes whose open ends are attached to the nine-branes.
By compactifying further on $S_1$ and interchanging the two compact directions,
the heterotic string theory turns into S-dual, Type I string theory.
Equally interesting situation is when M-theory is compactified on orbifold
$({\bf S}_1)^5/{\bf Z}_2$ with 16 five-branes at fixed
points~\cite{wittensworks}.
Townsend~\cite{townsend} and Strominger~\cite{strominger} have shown that
in M-theory two-branes can also end on five-branes. In this case, the
boundary of open two-brane is a self-dual string propagating on the
five-brane world-volume. If the separation between the two ends shrinks,
the tension of self-dual string decreases and a non-critical tensionless
string emerges.

If the M(atrix) theory provides a non-perturbative definition of M-theory,
the situations alluded above point to a problem of formulating
M(atrix) theory on an orbifold and in the presence of five- and/or
nine-branes. In particular, M(atrix) theory should support twisted
two-branes with open ends attached to these branes.
An interesting and important question is M(atrix) theory explanation
to the origin and the enhancement of gauge symmetries from the twisted sector.
In this regard, M(atrix) theory defined on an orbifold may
shed new non-perturbative understanding to the dynamics of $D=10$ strongly
coupled heterotic string and of $D=6$ non-critical self-dual string.

In this paper, with these motivations, we investigage M(atrix) theory
defined on an orbifold ${\bf S}_1/{\bf Z}_2$ with particular emphasis on
the strongly coupled heterotic string theory. As such, we will call
it as heterotic M(atrix) theory.
In section 2, we analyze Chan-Paton factors of open strings
connecting 0-brane partons. We show that the orbifold conditions that reverse
both the spacetime parity and the string orientation allow
$SO(2N), SO(2N+1)$ and $USp(2N)$ as possible gauge groups of the M(atrix)
theory. It is also necessary, however, to check if appropriate two-branes
in the twisted sector can arise from a chosen gauge group.
In section 3, we find that possible topologies of two-brane are intimately
related to the choice of M(atrix) theory gauge group.
Analysis of area-preserving diffeomorphism gauge symmetry indicates
that only $SO(2N)$ and $SO(2N+1)$ groups are capable of describing both
open and closed two-branes, while $USp(2N)$ group describes
only closed two-branes. In section 4, we construct heterotic M(atrix) theory.
We identify the twisted sector spectra that cancels {\sl locally } both
one-loop vacuum energy and worldsheet anomaly of resulting heterotic string.
We also find macroscopic M(atrix) open and closed two-branes and derive
correspondence rules in the large N limit.

Our notation of M(atrix) theory is as follows.
Regularizing zero-momentum limit \footnote{It should
be noted that non-trivial vacuum phenomena such as axial anomaly requires
careful treatment of zero-momentum components\cite{jirey} in the infinite
momentum frame.} by compactifying the
longitudinal direction on a circle of radius $R$, the M(atrix) theory
action is given by a matrix quantum mechanics of $SU(N)$ gauge group
\cite{banksetal} :
\be
S_M = {\rm Tr}_N \int \! d\tau \, \Big(
{1 \over 2 R} (D_\tau {\bf X}^I)^2 + {R \over 4} [{\bf X}^I , {\bf X}^J]^2
+ {\bf \Theta}^T D_\tau {\bf \Theta} + i R {\bf \Theta}^T \Gamma_I [{\bf X}^I ,
{\bf \Theta}] \Big).
\label{action}
\ee
Here, ${\bf X}^I$ and ${\bf \Theta}^\alpha$ denote 9 bosonic and 16 spinor
coordinates of 0-brane partons ($I = 1, \cdots, 9$ and $\alpha = 1, \cdots,
16$). The Majorana spinor conventions are such that $\Gamma_I$'s are real
and
symmetric and $i {\overline {\bf \Theta}} \Gamma_- \equiv {\bf \Theta}^T$:
\be
\Gamma_i = \left( \begin{array}{cc}
0 & \sigma^i_{a {\dot a}} \\ \sigma^i_{{\dot a} a} & 0 \end{array} \right)
\hskip0.5cm i = 1, \cdots, 8; \hskip1cm
\Gamma_9 = \left( \begin{array}{cc}
- \delta_{a b} & 0 \\ 0 & + \delta_{{\dot a} {\dot b}} \end{array} \right).
\label{gammamatrix}
\ee
The non-dynamical gauge field $A_\tau$ that enters through covariant
derivatives $D_\tau {\bf X}^I \equiv \partial_\tau {\bf X}^I -
i [A_\tau, {\bf X}^I]$ and
$D_\tau {\bf  \Theta}^\alpha \equiv \partial_\tau {\bf \Theta}^\alpha -
i [A_\tau, {\bf \Theta}^\alpha]$ projects the physical Hilbert space to
a gauge singlet sector and ensures invariance under area-preserving
diffeomorphism transformation. Thus, M(atrix) theory consists of
 $9 (N^2 - 1) - (N^2 - 1)$
bosonic degrees of freedom and $16 (N^2 - 1)/2 = 8 (N^2 - 1)$ spinor degrees
of freedom.
Hamiltonian in the infinite momentum limit is given by
\be
H_M = R \, {\rm Tr}_N \Big( {1 \over 2} {\bf \Pi}^2_I
- {1 \over 4} [{\bf X}^I, {\bf X}^J]^2 + i {\bf \Theta}^T \Gamma_I [{\bf X}^I,
{\bf \Theta}] \Big).
\label{hamiltonian}
\ee
The M(atrix) theory is invariant under the following supersymmetry
transformations
\bee
\delta {\bf X}^I &=& - 2 \epsilon^T \Gamma^I {\bf \Theta} \nonumber \\
\delta {\bf \Theta} &=& { i \over 2} \Big( \Gamma_I D_\tau {\bf X}^I
+ {1 \over 2} \Gamma_{IJ} [{\bf X}^I, {\bf X}^J] \Big) \, \epsilon + \xi
\nonumber \\
\delta A_\tau &=& - 2 \epsilon^T \, {\bf \Theta}.
\label{susytransformation}
\eee
Here, $i \epsilon$ is a 16-component spinor generator of local
supersymmetry, while $\xi$ is a 16-component spinor generator of rigid
translation\footnote{Invariance under the kinematical supersymmetry
can be checked easily using cyclic property of the trace operation.}.
As such, the sixteen dynamical and sixteen kinematical supersymmetry charges
are given by:
\bee
{\bf Q}_\alpha &=& {\sqrt R} {\rm Tr} \Big( \Gamma^I {\bf \Pi}_I +
{i  \over 2} \Gamma_{IJ} [{\bf X}^I, {\bf X}^J] \Big)_{\alpha \beta}
{\bf \Theta}_\beta,
\nonumber \\
{\bf S}_\alpha &=& {2 \over \sqrt R} {\rm Tr} {\bf \Theta}_\alpha
\label{susycharges}
\eee
respectively.

\section{Chan-Paton Factor Analysis}
\setcounter{equation}{0}

We first determine what gauge groups are capable of describing 0-brane
parton interactions on an orbifold and in the presence of spatial boundaries.
In the next section we will examine compatibility with the area-preserving
diffeomorphism invariance and topology of classical two-branes
that arise as a collective excitation in the respective M(atrix) theory.
We first try to describe the group theory aspects
which arise when we consider orbifolds in M(atrix) theory.
As we will see, the allowed M(atrix) theory gauge groups are $SO(2N),
SO(2N+1)$ or $USp(2N)$.

A convenient way of describing an orbifold is in terms of covering
space. The effect of orbifold is then to put both original and `image'
0-brane partons together in the covering space and impose appropriate
projection condition. This projection is a proper sub-group of the original
M(atrix) theory. Therefore, in this description, coordinate and spinor
matrices are $2N \times 2N$ in size.
Denoting coordinate matrices parallel and perpendicular
to a boundary by ${\bf X}_{\parallel}$ and ${\bf X}_\perp$ respectively,
the Chan-Paton conditions to coordinate and spinor matrices are given by:
\bee
{\bf X}_{\perp} &=& - M \cdot {\bf X}^{\rm T}_{\perp} \cdot M^{-1},
\nonumber \\
{\bf X}_{\parallel} &=& + M \cdot {\bf X}^{\rm T}_{\parallel} \cdot M^{-1},
\nonumber \\
{\bf \Theta} &=& \Gamma_\perp M \cdot {\bf \Theta}^{\rm T} \cdot M^{-1},
\hskip1cm \Gamma_\perp \equiv \prod_{I \in \perp} \Gamma_I.
\label{refl}
\eee
(The same boundary conditions have been observed previously \cite{ferretti,
motl} but with a different interpreation or in a limited generality.)
Here, $\Gamma_\perp$ denotes an ordered product of perpendicular direction
gamma matrices. Note that $\{ \Gamma_\perp, \Gamma_{\parallel} \} = 0$
for any parallel direction gamma matrices.

One can easily check that the orbifold mapping Eq.(\ref{refl})
is a symmetry of the M(atrix) theory
Hamiltonian Eq.(\ref{hamiltonian}), hence, allows an orbifold projection.
The symmetry is related to the type IIA symmetry that reverses the spacetime
parity and the string orientation simultaneously. In M-theory, the
corresponding symmetry is a simultaneous spatial reflection and reversal of
antisymmetric 3-form tensor potential $A_{MNP} \rightarrow - A_{MNP}$.

The matrix $M$ relates the original and the `image'
0-brane partons. Hermiticity of the coordinate and spinor
matrices requires that $M^{\rm T} \cdot M^{-1} = \pm {\bf I}_{2N \times 2N}$,
viz. $M$ is either a symmetric or an anti-symmetric matrix.

It is straightforward to check that the $\pm$ signs in the parallel and the
perpendicular coordinates manifest orbifold projections: orignal and `image'
0-brane should have opposite perpendicular coordinates but the same parallel
coordinates. To see this, consider a characteristic polynomial $P(x)$ of
eigenvalues of ${\bf X}$:
\bee
P(x) \equiv
{\rm det} (x {\bf I} - {\bf X}) & = &
{\rm det} (x {\bf I} \pm M \cdot {\bf X}^{\rm T} \cdot M^{-1}) \nonumber \\
& = &
{\rm det} (\mp x {\bf I} - {\bf X}) = P(\mp x)
\label{coordinatepairing}
\eee
for perpendicular and transverse coordinates respectively.
Note that the result depends only on the overall sign and independent of
the matrix $M$.

The matrix $M$ imposes one-to-one relation between an original parton and
its `image' parton. We have shown that $M$ is either symmetric or
anti-symmetric matrix. To analyze the resulting gauge group of M(atrix) theory
for each possible choices of $M$, it is convenient to consider ${\bf X}$ as
an assemble of 4 blocks $N\times N$ matrices. The possible choices of $M$ can
be written as:
\be
(M_\mu)_{2N \times 2N} = {\bf I}_{N \times N} \otimes \sigma_\mu
\ee
where $\sigma_\mu \equiv (1, \sigma_1, \sigma_2, \sigma_3)$ is a complete set
of basis of $2 \times 2$ matrices.
Symmetric and anti-symmetric choices of $M$ correspond to
$M_0, M_1, M_3$ and $M_2$ respectively.
For $M = M_1$ choice, one finds that
\be
{\bf X}_\perp = \left( \begin{array}{cc}
Y & A \\ - A^* & - Y^T \end{array} \right)
\ee
where $Y$ is a Hermitian matrix and $A$ is an anti-symmetric, complex matrix.
The total number of independent elements is $N^2 + 2 \cdot N(N-1)/2 =
2N(2N-1)/2$. On the other hand, the parallel coordinate matrices take a form of
\be
{\bf X}_{\parallel} = \left( \begin{array}{cc}
Z & S \\ S^* & Z^T \end{array} \right)
\ee
where $Z$ is an unconstrained Hermitian matrix and $S$ is a symmetric, complex
matrix. The total number of independent elements is $N^2 + 2 \cdot
N(N+1)/2 = 2N (2N + 1)/2$.

By unitary transformation, the above matrices can be brought into
anti-symmetric and symmetric, Hermitian matrices. Consider
${\bf X} \rightarrow U \cdot {\bf X} \cdot U^\dagger$ where
\be
U = {1 \over \sqrt 2} \left( \begin{array}{cc}
                      1 & i \\ i & 1 \end{array} \right)
\ee
Then,
\bee
{\bf X}_\perp &\rightarrow & {1 \over 2}
\left( \begin{array}{cc}
(Y - Y^{\rm T}) - i (A + A^*) & - i (Y + Y^{\rm T}) + (A - A^*) \\
i (Y + Y^{\rm T}) + (A - A^*) & (Y - Y^{\rm T} ) + i (A + A^*)
\end{array} \right)
\nonumber \\
{\bf X}_{\parallel} &\rightarrow& {1 \over 2}
\left( \begin{array}{cc}
(Z + Z^{\rm T}) - i (S - S^*) & - i (Z - Z^{\rm T} ) + (S + S^*) \\
+ i (Z - Z^{\rm T}) + (S + S^*) & (Z + Z^{\rm T}) + i (S - S^*)
\end{array} \right).
\eee
It clearly displays that ${\bf X}_\perp$ and ${\bf X}_{\parallel}$ are
Hermitian, anti-symmetric and symmetric matrices. As such, they form
adjoint and symmetric representations of $SO(2N)$ respectively.

Other symmetric choices $M = M_0, M_3$ give similar results: while
block structure depends on a specific choice of $M$, the perpendicular and
the parallel coordinate matrices are matrices having $2N (2N-1)/2$ and
$2N(2N+1)/2$ independent elements respectively.
They are precisely adjoint and symmetric
representations of $SO(2N)$.
We conclude that the M(atrix) theory on an orbifold with
symmetric choice of $M$ is described by
a matrix quantum mechanics with gauge group $SO(2N)$.

Next, consider the choice of anti-symmetric matrix $M = M_2$.
In this case, one finds
that the perpendicular and the parallel coordinates take block matrix
structure of:
\bee
{\bf X}_\perp &=& \left( \begin{array}{cc}
Y & S \\ S^* & - Y^{\rm T} \end{array} \right),    \nonumber \\
{\bf X}_{\parallel} &=& \left( \begin{array}{cc}
Z & A \\ - A^* & Z^{\rm T} \end{array} \right).
\label{uspmatrices}
\eee
In this case, the perpendicular coordinate has $N^2 + 2 \cdot N(N + 1)/2
= 2N(2N+1)/2$ independent elements, while the parallel coordinates have
$N^2 + 2 \cdot N(N-1)/2 = 2N (2N-1)/2$ independent elements. As such, they
may be identified with adjoint and antisymmetric representations of $USp(2N)$
group.
In fact, with $M_2 = {\bf I}_{N \times N} \otimes \sigma_2$, the orbifold
condition for the transverse coordinate
\be
\left( \begin{array}{cc} A & B \\ C & D \end{array} \right)
\left( \begin{array}{cc} 0 & - 1 \\ + 1 & 0 \end{array} \right)
+
\left( \begin{array}{cc} 0 & + 1 \\ - 1 & 0 \end{array} \right)
\left( \begin{array}{cc} A^{\rm T} & C^{\rm T} \\ B^{\rm T} & D^{\rm T}
\end{array} \right)
= 0
\ee
is nothing but the condition of $Sp(2N, {\bf C})$ for $SU(2N)$ Hermitian
matrices.
We conclude that M(atrix) theory on a space with boundaries with an
anti-symmetric choice of $M$ is described a by matrix quantum mechanics with
gauge group $USp(2N)$.

Finally, we analyze the fermion coordinate matrices. Due to the light-cone
projection $\Gamma_+ {\bf \Theta} = 0$, there are only 16-components. They
form a spinor representation of $SO(9)$ of the transverse space.
Consider a pair of longitudinal nine-branes ("end of the world") located
perpendicular to the 9-th direction, as is relevant for ${\bf S}_1/{\bf Z}_2$.
The residual transverse space manifests $SO(8)$ rotational symmetry.
The 16-component spinor can be further decomposed into two inequivalent chiral
spinors of $SO(8)$. Thus,
\be
{\bf \Theta} = {\bf 8}_s \oplus {\bf 8}_c \equiv {\bf S}_a \oplus
{\bf S}_{\dot a}
\label{spinordecomposition}
\ee
where
\be
\Gamma_9 {\bf S}_a = - {\bf S}_a; \hskip0.5cm
\Gamma_9 {\bf S}_{\dot a} = + {\bf S}_{\dot a}.
\label{chirality}
\ee
Under the orbifold condition, for gauge group $SO(2N)$, ${\bf S}_a ,
{\bf S}_{\dot a}$ are adjoint representation and symmetric representations
respectively. For the gauge group $USp(2N)$,
${\bf S}_a$ is an adjoint representation while
${\bf S}_{\dot a}$ is an anti-symmetric representation. Altogether,
the 16-components of the spinors ${\bf 8}_s \oplus
{\bf 8}_c$ have total $16 N^2$ on-shell degrees of freedom for both
$SO(2N)$ and $USp(2N)$ gauge groups. Compared to the total bosonic on-shell
degrees of freedom $2N(2N \mp 1)/2 + 8 \cdot 2N(2N \pm 1)/2 - 2N(2N \mp 1)/2$,
there are $16N$ fermionic or $8N$ bosonic degrees of freedom deficit for
$SO(2N)$ and $USp(2N)$ gauge groups respectively.
It should be noted that center of mass degrees of freedom consist of
8 bosonic and 8 fermionic modes and
the deficit degrees of freedom originate from the internal relative
motion among the 0-brane partons.

The mismatch of Bose-Fermi degrees of freedom is consistent with
the surviving supersymmetries of the orbifold projection and $SO(2N)$ or
$USp(2N)$ gauge symmetries.
As will be discussed in section 4, however, mismatch of the degrees of
freedom leads to non-trivial one-loop renormalization of the moduli space
metric and linear potential for the 0-brane.
The radiatively induced potential amounts to dilaton tadpole amplitude, hence,
local cosmological constant. Such a potential can be cancelled {\sl locally}
by introducing a twisted sector of the orbifold.
From the above counting, we see that the twisted sector should consist of
8 fermions (4 bosons) in the fundamental
representation of $SO(2N)$ ($USp(2N)$) gauge group.
For now, it is sufficient to notice that the deficit degrees of freedom
originates from the orbifold conditions that eliminates $8N$ fermionic
or bosonic 0-brane partons excessively along the diagonal entries.

As a final remark, it is also possible to have 0-brane partons attached
to the orbifolds. They are interpreted as a mirror pair of `half' 0-brane
partons moved to the orbifold fixed point. Clearly such a situation is
described by $SO(2N+1)$ M(atrix) theory uniquely.

\section{Area-Preserving Diffeomorphism Analysis}
\setcounter{equation}{0}

It has been shown that the M(atrix) theory contains two-brane as a
collective excitation of Landau-level orbiting 0-brane partons forming
an incompressible two-dimensional fluid. The existence of two-brane
may be traced down to the area-preserving diffeomorphism invariance of
the M(atrix) theory in the $N \rightarrow \infty$ limit. In other
words, the M(atrix) theory is a gauge theory of area-preserving
diffeomorphism group~\cite{dewit}.

In the previous section, we have identified $SO(2N)$, $SO(2N+1)$ and
$USp(2N)$ as possible gauge groups of the M(atrix) theory on an orbifold.
Several possible choices of M(atrix) theory gauge
groups stems from the fact that the Chan-Paton factor analysis was purely
kinematical. If M(atrix) theory based on these gauge groups were to
describe open two-branes attached at the orbifold fixed points, then the
M(atrix) theory should possess area-preserving diffeomorphism invariance
relevant to their topologies. It is clear that
this requirement is more stringent than the Chan-Paton factor analysis,
hence, renders more refined information for the choice of M(atrix)
theory gauge group.

In this section, with the above motivation, we analyze the area-preserving
diffeomorphism invariance relevant to two-branes of lower genus topologies.
It is well-known that M(atrix) theory with $SU(N)$
group is a gauge theory of area-preserving diffeomorphism for both spherical
and toroidal two-branes, ${\rm Diff}_0 (S_2)$ and ${\rm Diff}_0 (T_2)$.
The area-preserving
diffeomorphism of open two-branes of disk $D_2$ and cylinder $C_2$
topologies are then obtained by truncating generators of
${\rm Diff}_0 (S_2)$ or ${\rm Diff}_0 (T_2)$ appropriately.
In fact, all other topologies
related to $S_2$ and $T_2$ by involution, M\"obius strip $M_2$,
Klein bottle $K_2$ and real projective space $RP_2$ are also obtained in
this manner. Some of these unoriented and closed topologies have been studied
previously by Pope and Romans\cite{poperomans} and
by Fairlie, Fletcher and Zachos\cite{
zachosetal}.
For completeness, we include their results briefly in the foregoing list.

As will be shown below, $SO(2N)$ M(atrix) theory describes disk $D_2$,
cylinder $C_2$, Klein bottle $K_2$, and M\"obius strip $M_2$. $SO(2N+1)$
M(atrix) theory describes disk $D_2$, cylinder $C_2$ and M\"obius strip $M_2$,
hence, open branes only. On the other hand, $USp(2N)$ M(atrix) theory
yields real-projective space $RP_2$, and Klein bottle $K_2$, hence,
unoriented closed two-branes only (The result of our analysis is summarized
in the table at the end of this section). We thus conclude that only
$SO(2N)$ and $SO(2N+1)$ M(atrix) theories are capable of having open
two-branes.
\subsection{Genus 0 and 1/2 Two-Branes}
By involution of spherical two-brane with ${\rm Diff}_0 (S_2)$, it is possible
to construct disk $D_2$ and real-projective plane $RP_2$ two-branes.
We now show that
the M(atrix) theory describing these two-branes, hence, preserving
${\rm Diff}_0 (D_2)$ and ${\rm Diff}_0 (RP_2)$ area-preserving
diffeomorphism invariance is given by $SO(2N)$, $SO(2N+1)$ or
$USp(2N)$ gauge groups respectively.
\subsubsection{Spherical Two-Brane and ${\rm Diff}_0 (S_2)$}
To describe area-preserving diffeomorphism of sphere $S_2$, following
Ref.\cite{poperomans}, let us introduce a complete set of scalar spherical
harmonics
\be
Y_{\bf lm} ({\bf x}) \equiv
C_{i_1, \cdots, i_l} x^{i_1} \cdots x^{i_l}
\label{sphericalharmonics}
\ee
in terms of embedding space coordinates ${\bf x} \equiv
(x_1, x_2, x_3)$ satisfying
\be
{\bf x} \cdot {\bf x} = 1.
\ee
In Eq.(\ref{sphericalharmonics}), $C_{i_1, \cdots, i_n}$ are symmetric,
traceless, tensor coefficients. The `magnetic' quantum number
ranges over $\bf m$: $ - {\bf l} \le {\bf m} \le  {\bf l}$.
The $S_2$ area-preserving diffeomorphism
algebra is encoded into the Poisson bracket algebra among the spherical
harmonics
\bee
\{ Y_{\bf lm}, , Y_{{\bf l'm'}} \} & \equiv & \epsilon_{ijk} x^i
(\partial_j Y_{\bf lm}) ( \partial_k Y_{{\bf l'm'}})
\nonumber \\
&=& \bigoplus_{{\bf j} = |{\bf l} - {\bf l}'|}^{{\bf l} + {\bf l}' - 1}
[\, Y_{\bf j(m+m')} \, ],
\label{bracket}
\eee
viz. a sum of irreducible polynomials of scalar harmonics in the range
$|{\bf l} - {\bf l}' | \le {\bf j} \le ({\bf l} + {\bf l}' - 1)$.

The above construction of area-preserving diffeomorphism algebra is in one-to-one
correspondence with $SU(N)$ Lie algebra expressed in terms of maximal
embedding of $SU(2)$. Under maximal embedding,
the generators of $SU(N)$ can be expressed as products of $SU(2)$
generators $\Sigma_i$ in the $N$-dimensional representation:
\bee
T^{(1)} &=& C_i \Sigma_i
\nonumber \\
T^{(2)} &=& C_{ij} \Sigma_i \Sigma_j \nonumber \\
\cdots
\nonumber \\
T^{(N-1)} &=& C_{i_1 i_2 \cdots i_{N-1}} \Sigma_{i_1} \cdots
\Sigma_{i_{N-1}}.
\label{sungenerators}
\eee
Here, the coefficients $C_{ijk \cdots}$ are the same symmetric, traceless,
tensor coefficients as in Eq.(\ref{sphericalharmonics}).
The above form of $SU(N)$ generators simply express that,
using the fact that the fundamental $N$-dimensional
representation of $SU(N)$ remains irreducible in $SU(2)$, the adjoint
representation of $SU(N)$ is decomposed into $N^2 - 1 =
3 + 5 + \cdots + (2N-1)$ representations of $SU(2)$.
The $T^{(i)}$ matrices form a complete set of traceless, Hermitian
$N \times N$ matrices, hence, provide a basis for $SU(N)$.

The $\Sigma_i$ generators in the $N$-dimensional representation of $SU(2)$
can be represented by a totally symmetrized $2^{(N-1)} \times
2^{(N-1)}$ matrices:
\be
\Sigma_i = {\rm Sym} [
\sigma_i \otimes {\bf I} \otimes \cdots \otimes {\bf I}+
{\bf I} \otimes \sigma_i \otimes {\bf I} \otimes \cdots \otimes {\bf I}
+\cdots+{\bf I} \otimes {\bf I} \otimes \cdots \otimes \sigma_i], \hskip0.5cm
i=1,2,3
\label{su2generators}
\ee
in which $\sigma_i, i = 1,2,3$ are the Pauli matrices.

Comparison of Eq.(\ref{sphericalharmonics}) and Eq.(\ref{sungenerators})
shows that the commutation relation among $SU(N)$ generators $T_{(i)}$
is isomorphic to the Poisson bracket relation Eq.(\ref{bracket}) among the
spherical harmonics $Y_{\bf l}({\bf x})$. In the large $N$ limit,
the non-commutativity among $T^{(i)}$'s becomes irrelevant. Thus,
the area-preserving diffeomorphism of $S_2$, ${\rm Diff}_0 (S_2)$ is
realized by $SU(N)$ algebra, reproducing well-known
result\cite{dewit2}.

\subsubsection{$R/{\bf Z}_2$ Two-Brane and ${\rm Diff}_0 (S_2)$}
Two-brane attached on a boundary at fixed point of $R / {\bf Z}_2$ is
topologically equivalent to disk $D_2$ two-brane. The latter is obtained
by an involution
\be
x^3 \rightarrow - x^3
\label{involution}
\ee
from the $S_2$ two-brane constructed above.
In terms of polar angle $(\theta, \phi)$, the involution Eq.(\ref{involution})
is equivalent to simultaneous parity and
$\phi \rightarrow \phi + \pi$ rotation, under which
\be
Y_{\bf lm} ({\bf x}) \rightarrow (-)^{\bf l + m} Y_{\bf l m} ({\bf x}).
\label{polarinvolution}
\ee
The vector harmonics that form a basis of generators of
area-preserving diffeomorphism
transformation are obtained as parity-odd combinations:
\be
L_{\bf lm} \equiv \{ Y_{\bf lm}: Y_{\bf lm} - (-)^{\bf l+m} Y_{\bf lm} \}.
\label{d2subset}
\ee
Since only odd values of $\bf
(l+m)$ are selected as the basis, the total number
of vector harmonics generators are given by
\be
L_{\bf lm} = \{ Y_{1,0}, \, Y_{2,+1}, \, Y_{2,-1}, \, Y_{3,+2} , \, Y_{3,0},
\, Y_{3, -2} , \cdots \},
\ee
hence, yield $1 + 2 + 3 + \cdots + (2N - 1) = 2 N ( 2N - 1) / 2$ generators.
This equals precisely to the number of generators of $SO(2N)$ group, and
suggests that area-preserving diffeomorphism on $D_2$ is described by $N
\rightarrow \infty$ limit of $SO(2N)$ Lie algebra \footnote{
A similar analysis shows that $SO(2N+1)$ subgroup is also possible by recalling
that $1 + 2 + \cdots + (N-1) = N(N-1)/2$. }.

We now show that the above construction of $D_2$ area-preserving
diffeomorphism is in one-to-one correspondence with the Lie algebra of
$SO(N)$. Again, using the maximal embedding of $SU(2)$ in $SU(N)$ and
corresponding representation of the basis Eq.(\ref{sungenerators}),
it remains to show that the generators are Hermitian and antisymmetric.
The symmetric tensor $C_{i_1 \cdots i_n}$ relevant for vector harmonics
satisfying the involution Eq.(\ref{involution}) allows only odd numbers of
$\Sigma_3$. It is now convenient to make $(\Sigma_1, \Sigma_2, \Sigma_3)_{D_2}
= (\Sigma_3, \Sigma_1, \Sigma_2)_{S_2}$. This cyclic permutation
of $N$-dimensional $SU(2)$ generators redefines $\Sigma_3$ naturally into an
anti-symmetric matrix :
\be
\Sigma_3 = {\rm Sym} [ \sigma_2 \otimes {\bf I} \otimes \cdots \otimes {\bf I}
+ {\bf I} \otimes \sigma_2 \otimes {\bf I} \otimes \cdots \otimes {\bf I}
+ {\bf I} \otimes {\bf I} \otimes \cdots \otimes \sigma_2].
\ee

Noting that the constant tensor $C_{ijk\cdots}$'s
are completely symmetric, we find that only a set of generators left over
are $2N(2N-1)/2$ independent, $2N \times 2N$ Hermitian, anti-symmetric
matrices. They are the generators of $SO(2N)$.

Again, in the large $N$ limit, the non-commutativity among the surviving
$T_{(i)}$'s die off sufficiently fast that the resulting $SO(N)$ Lie algebra
is exactly the same as the area-preserving diffeomorphism algebra
${\rm Diff}_0 (D_2)$.

\subsubsection{Real-Projective Two-Brane and ${\rm Diff}_0 (RP_2)$}
The real-projective plane is described by an involution ${\bf x}
\rightarrow - {\bf x}$ of the sphere $S_2$. Under the involution the
spherical harmonics maps as $Y_{\bf lm} \rightarrow (-)^{\bf l}
Y_{\bf lm}$. Hence, a complete set of vector
harmonics that generate the area-preserving diffeomorphism ${\rm Diff}_0
(RP_2)$ are the odd-parity subset of $S_2$ spherical harmonics
Eq.(\ref{sphericalharmonics}):
\be
L_{\bf lm} = \{  Y_1, \,\, Y_3 , \,\, Y_5 , \, \cdots, \, Y_{2N-1} \}
\ee
Hence, the Poisson algebra among these subset of harmonics is isomorphic
to sub-algebra that closes among the generators:
\be
T^{(1)}, \,\,T^{(3)}, \,\,T^{(5)}, \cdots, T^{(2N-1)}.
\label{subset}
\ee

As shown by Pope and Romans\cite{poperomans}, this sub-algebra
forms $Sp(2N, {\bf C}) \, \cap \, SU(2N) = USp(2N)$ group.
To see this,
consider a totally anti-symmetric $(2^{(2N-1)} \times 2^{(2N-1)})$ matrix
${\cal M}$:
\be
{\cal M} \equiv {\rm Sym} [ \sigma_2 \otimes \sigma_2 \otimes \cdots
\otimes \sigma_2].
\label{symplecticmetric}
\ee
It is straightforward to check that the $SU(2)$ generators $\Sigma_i$
in Eq.(\ref{su2generators}) of the $2N$-dimensional representation
satisfies:
\be
\Sigma_i \cdot {\cal M}  + {\cal M} \cdot \Sigma_i^{\rm T} = 0,
\ee
hence, for $i = 1, 3, \cdots, (2N-1)$,
\be
T_{(i)} \cdot {\cal M} + {\cal M} \cdot T_{(i)}^{\rm T} = 0.
\ee
Therefore, the subset of generators Eq.(\ref{subset}) forms an
$Sp(2N, {\bf C})$ algebra. Since they are Hermitian as well, the
generators actually closes under $Sp(2N, {\bf C}) \cap SU(2N) = USp(2N)$.
This establishes that M(atrix) theory with a gauge group $USp(2N)$
is relevant to real-projective plan two-brane and ${\rm Diff}_0 (RP_2)$
thereof.

\subsection{Genus-1 Two-Branes}
Starting from toroidal two-brane, open two-branes of cylindrical and
M\"obius strip topologies, and closed and unoriented
two-brane of Klein-bottle topology
are obtained by involution.
\subsubsection{Toroidal Two-Brane and ${\rm Diff}_0 (T_2)$}
Consider a torus defined in terms of coordinates $(x,y)$
\be
{\bf x} \equiv (x,y) \simeq (x + 2 \pi, y) \simeq (x , y + 2 \pi).
\label{domain}
\ee
A complete set of scalar harmonics on $T_2$ are Fourier modes:
\be
Y_{\bf m} ({\bf x}) \equiv \exp [i {\bf m} \cdot {\bf x}];
\hskip1cm {\bf m} = (m, n) \in {\bf Z^2}.
\label{t2harmonics}
\ee
The area-preserving diffeomorphisms are generated by vector fields
\bee
T_{\bf m} &=& - (\partial_a Y_{\bf m}) \epsilon^{ab} \partial_b
\nonumber \\
&=&
Y_{\bf m} {\bf m} \times (-i \nabla_{\bf x}).
\label{generators}
\eee
They satisfy ${\rm Diff}_0 (T_2)$ algebra
\be
[T_{\bf m} , T_{\bf n}] = ({\bf m} \times {\bf n}) T_{\bf m + n}.
\label{difft2algebra}
\ee

As is well-known\cite{dewit2, zachosetal}, finite matrix algebra realizes the above
${\rm Diff}_0 (T_2)$ in the $N \rightarrow \infty$ limit.
Specifically, one first picks a primitive $N$-th root of unity $\omega
= \exp (2 \pi i / N)$ and, following 't Hooft, a pair of unitary and traceless
matrices $U, V$ that are primitive $N$-th root of unity:
\be
UV = \omega V U; \hskip0.5cm U^N = V^N = 1.
\label{uvrelation}
\ee
It then follows that a set of matrices
\be
J_{\bf m} \equiv \omega^{-mn/2} U^m V^n
\label{matrixbasis}
\ee
for $1 \le m, n \le N$ are linearly independent, complete set of basis for
the $N \times N$ Hermitian matrices of $SU(N)$ (excluding the rigid mode
generator $J_{00}$).
From the composition rule:
\be
J_{\bf m} J_{\bf n} =
\omega^{{\bf m} \times {\bf n}/2} J_{\bf m + n},
\ee
the commutators of ${\bf T}_{\bf m} \equiv  N / 2 \pi i J_{\bf m}
$ are closed:
\be
[{\bf T}_{\bf m}, {\bf T}_{\bf n}]
= {N \over \pi} \sin \Big({\pi \over N} {\bf m} \times {\bf n}\Big)
{\bf T}_{\bf m + n}.
\label{matrixcommutator}
\ee
In the large N limit, the commutator algebra Eq.(\ref{matrixcommutator})
agrees with Eq.(\ref{difft2algebra}),
the area-preserving diffeomorphism algebra ${\rm Diff}_0 (T_2)$.
Correspondingly, M(atrix) theory with gauge group $SU(N)$ possesses
${\rm Diff}_0 (T_2)$ and describes $T_2$ two-branes as collective excitations
of 0-brane partons.


\subsubsection{$S_1/{\bf Z}_2$ Two-Branes and ${\rm Diff}_0 (C_2)$}
Cylindrical two-brane is obtained by an involution of $T_2$:
\bee
(x,y) \simeq (x,2\pi-y),
\label{c2involution}
\eee
so that the fundamental cell $[0, 2\pi] \times [ 0 , \pi ]$ is
double-covered by $T_2$.
Under the involution, the scalar harmonics transforms as
\be
Y_{\bf m} \rightarrow  Y_{\tilde {\bf m}}; \hskip1cm
{\tilde {\bf m}} \equiv (m, - n).
\label{scalarharmonics}
\ee
Therefore, the vector harmonics of the area-preserving diffeomorphism
on $C_2$ is spanned by the subset of basis:
\be
C_{\bf m} \equiv \{ Y_{\bf m} : Y_{\bf m} -  Y_{\tilde {\bf m}} \}.
\label{c2basis}
\ee
They satisfy the ${\rm Diff}_0 (C_2)$ algebra:
\be
[ C_{\bf k} ,  C_{\bf m} ] = ({\bf k} \times {\bf m}) C_{\bf k + m}
-  ({\bf k} \times {\tilde {\bf m}}) C_{{\bf k} + {\tilde {\bf m}}}.
\label{c2diffalgebra}
\ee

We now realize the above ${\rm Diff}_0 (C_2)$ algebra in terms of large
$N$ limit matrix algebra. Using the commutation relation
Eq.(\ref{matrixcommutator}), it is easy to show that the set of matrices
\be
{\bf C}_{mn} \equiv { N \over 2 \pi i} [ J_{\bf m} - J_{\tilde {\bf m}} ]
\ee
are closed and form a subalgebra of $SU(N)$. Let us take
$U$ and $V$ as
\be
U =
\left(
\begin{array}{ccccc}
\omega &&&& \\ & \omega^2 &&& \\ && \omega^3 && \\ &&& \ddots & \\
&&&& 1
\end{array}
\right),
\hskip0.5cm
V =
\left(
\begin{array}{ccccc}
&&&& 1 \\ 1 &&&& \\ & 1 &&& \\ && \ddots && \\ &&& 1 &
\end{array}
\right),
\ee
which have the property as $U^T=U$ and $V^T=V^{-1}$.  We find that
$ J^T_{\bf m} =  J_{\tilde {\bf m}} $
and matrices ${\bf C}_{\bf m}$ are anti-symmetric. Being Hermitian, they
span $SO(2N)$ or $SO(2N+1)$ Lie algebras.

Incidentally, there exists an alternative involution possible. Involution
of $T_2$ as:
\be
(x, y) \simeq (x, \pi - y)
\ee
yields the fundamental cell as $[0, 2 \pi] \times [\pi/2, 3\pi/2]$. In this
case, the generator matrices are projected to a subset consisting of
$J_{\bf m} - (-)^n J_{\tilde {\bf m}}$. They span $SO(2N) \in U(2N)$.

We conclude that M(atrix) theory defined with
a gauge group $SO(2N)$ or $SO(2N+1)$ possesses
${\rm Diff}_0 (C_2)$ local symmetry and
is capable of describing twisted two-branes of cylinder topology.
Precisely such a two-brane
stretched between the two ten-dimensional "end of the world" is expected
to describe strongly coupled heterotic and dual type IA strings.

\subsubsection{ M\"obius Strip Two-Brane and ${\rm Diff}_0 (M_2)$ }

The M\"obius strip is obtained from the torus by an involution:
\be
(x,y) \simeq (y,x).
\ee
The fundamental cell is a square made by 4 intersecting
lines, i.e. $y=\pm x$ and $y =\pm (x-\pi)$. $y=-x$ and $y=-x+\pi$
are glued together with a twist, thus giving a M\"obius strip.
Under this involution the scalar harmonics transform as:
\be
Y_{\bf m} \rightarrow  Y_{\hat {\bf m}}; \hskip1cm
{\hat {\bf m}} \equiv (n, m).
\ee
The parity-odd harmonics are generated by:
\be
M_{\bf m} = Y_{\bf m} -  Y_{\hat {\bf m}} .
\ee
They are closed under commutators:
\be
[M_{\bf k}, M_{\bf m} ] = ({\bf k} \times {\bf m}) M_{\bf k + m}
- ({\bf k} \times {\hat {\bf m}}) M_{\bf k + \hat {\bf m}}.
\ee
This basis form $SO(N)$, for N can be both even and odd.

Consider the matrices spanned by
\be
U = \omega^{-{N-1 \over 4}} \left(
\begin{array}{cccccc}
& \omega^{1/2} &&&& \\
&& \omega &&& \\
&&& \omega^{3/2} && \\
&&&& \ddots & \\
&&&&& \omega^{N-1 \over 2} \\
1 &&&&&
\end{array} \right),
\quad
V = U^T.
\ee
We have
\be
J_{\bf m} = J_{\bf \hat{m}}^T,
\ee
hence, the matrices $M_{\bf m}= J_{\bf m} - J^{\rm T}_{\hat {\bf m}}$
are antisymmetric. Being Hermitian at the same time, they span $SO(2N)$ and
$SO(2N+1)$ Lie algebra.
We conclude that M(atrix) theory with these groups possess
${\rm Diff}_0 (M_2)$ and is capable of describing twisted two-branes of
M\"obius strip topology.

\subsubsection{Klein-Bottle Two-Brane and ${\rm Diff}_0 (K_2)$ }
The Klein bottle $K_2$ is obtained from the torus $T_2$ by an involution of
simultaneous reflection and shift:
\be
(x,y) \simeq ( x + \pi, 2 \pi - y),
\label{k2involution}
\ee
viz. the fundamental domain $[0,  \pi] \times [0,  2 \pi]$ is double-covered
by $T_2$. Under the involution Eq.(\ref{k2involution}), the scalar
harmonics transforms as:
\be
Y_{\bf m} \rightarrow (-)^m Y_{\bf \tilde m};
\hskip1cm {\tilde {\bf m}} = (m, -n).
\label{harmonicsmap}
\ee
Therefore, the parity-odd vector harmonics are generated by a subset
of harmonics:
\be
K_{\bf m} =  L_{\bf m} - (-)^m L_{  \tilde {\bf m}}.
\label{k2vectorharmonics}
\ee
In addition, there is one global vector fields and associated rigid
generator. The harmonics $K_{\bf m}$ in Eq.(\ref{k2vectorharmonics})
are closed under commutators:
\be
[K_{\bf k}, K_{\bf m} ] = ({\bf k} \times {\bf m}) K_{\bf k + m}
- (-)^n ({\bf k} \times {\tilde {\bf m}}) K_{\bf k + \tilde {\bf m}}.
\label{k2commalgebra}
\ee

The matrix algebra realization of the above ${\rm Diff}_0 (K_2)$ is
proceeded by replacing $L_{\bf m}$ in the expression
Eq.(\ref{k2vectorharmonics}) by $(N/2 \pi i) J_{\bf m}$ of
Eq.(\ref{matrixbasis}). It turns out that the matrix algebra closes
only for even $N$. With the same choice for $U,V$ as was done for
cylinder, we have
\bee
MU^TM^{-1} &=& -U, \nonumber \\
MV^TM^{-1} &=& V^{-1}, \label{somaking}
\eee
where $M=\sigma_1 \otimes {\bf 1}$.
Using Eq.(\ref{somaking}), we have $MK^T_{\bf m} M^{-1}=-K_{\bf m}$.
We already know that with symmetric matrix $M$, above constraint
produce $SO(2N)$, as subalgebra of $U(2N)$. With slightly different choice,
we can also relate Klein bottle to $USp(2N)$. Take
\footnote{This choice satisfies $U^{2N}=V^{2N}=-1$, which is different
from usual convention.}
\be
U = \omega^{1/2} U_{\rm C_2}, \quad\quad
V =
\left(
\begin{array}{cccc|cccc}
&&&&&&& i \\ 1 &&&&&&&  \\
& \ddots &&&&&& \\ && 1 &&&&& \\ \hline
&&& i &&&& \\ &&&& 1 &&& \\
&&&&& \ddots && \\ &&&&&& 1 &
\end{array}
\right).
\ee
This choice satisfies Eq.(\ref{somaking}) again, with $M=\sigma_2 \otimes
{\bf 1}$. Since $M$ is antisymmetric now, we have reduction to $USp(2N)$.
Hence, both $SO(2N)$ and $USp(2N)$ can describe Klein-bottle two-brane.

To conclude, we summarize relation between the choice of M(atrix) theory
gauge group and twisted two-branes that can be described by each choice.

\vskip0.3cm
\begin{tabular}{|c|c|c|} \hline
M(atrix) Theory Gauge Group & Matrix Generators & Two-brane Topology \\ \hline
SO(2N) and SO(2N+1) & $Y_{\bf lm} - (-)^{\bf l + m} Y_{\bf lm}$
& Disk \\ \hline
USp(2N) & $Y_{\bf lm} - (-)^{\bf l} Y_{\bf lm}$ & Projective plane \\ \hline
SO(2N) and SO(2N+1) & $J_{\bf m} - J_{\bf {\tilde m}}$ & Cylinder \\ \hline
SO(2N) & $J_{\bf m}-(-)^nJ_{\bf {\tilde m}}$ & Cylinder  \\ \hline
SO(2N) and USp(2N) & $J_{\bf m}-(-)^mJ_{\bf {\tilde m}}$ & Klein bottle
\\ \hline
SO(2N) and USp(2N) & $J_{\bf m}-(-)^{m+n}J_{\bf {\tilde m}}$ & Klein bottle
 \\ \hline
SO(2N) and SO(2N+1) & $J_{\bf m}-J_{\hat {\bf m}}$ & M\"obius strip
\\ \hline
SO(2N) & $J_{\bf m} - (-)^{m+n} J_{\hat {\bf m}}$ & M\"obius strip \\
\hline
\end{tabular}
\vskip0.3cm

From this result, it should be clear that the twisted two-brane with open
ends at the 9-brane (end of the world) is possible only for the choice of
M(atrix) theory gauge group $SO(2N)$ and $SO(2N+1)$.

\section{Heterotic M(atrix) Theory}
\setcounter{equation}{0}

\subsection{Supersymmetry and Twisted Sector}
Based on the results of analysis on Chan-Paton factor and of area-preserving
diffeomorphism gauge symmetries, we have found that heterotic M(atrix) theory
is described by $SO(2N)$ or $SO(2N+1)$ matrix quantum mechanics\footnote{
SO(2N+1) matrix quantum mechanics was studied first in a closely related
context in Ref.~\cite{kachrusilverstein} based on earlier work Ref.~\cite{
ferretti}.}.
The thirty-two supersymmetries of the underlying $SU(2N)$ M(atrix) theory
is broken to sixteen supersymmetries due to the orbifold projections.
The field contents and gauge quantum numbers are given as follows.
For bosonic fields, 8 parallel coordinates ${\bf X}^i, \,\, i = 1, \cdots, 8$
are in the symmetric representations while transverse coordinate $A_9$ is in
the adjoint representation. The non-dynamical gauge field $A_0$ is also in
the adjoint representation. For spinor fields, eight component ${\bf S}_a$
are in the adjoint representations while eight component ${\bf S}_{\dot a}$
are in the symmetric representations. The surviving eight supersymmetries
are ${\bf 8}_s$ components. The heterotic M(atrix) theory Lagrangian is given
by:
\bee
L_{\rm untwisted} = {\rm Tr}  \Big( \!\!\!\!\!\! && \!\!\!\!\!\!
{1 \over 2 R} (D_\tau {\bf X}^i)^2
- {1 \over 2 R} (D_\tau A_9)^2 - {R \over 2} [A_9, {\bf X}^i]^2
+ {R \over 4} [{\bf X}^i, {\bf X}^j]^2 \nonumber \\
&-& {\bf S}_a D_\tau {\bf S}_a + R \, {\bf S}_a [A_9, {\bf S}_a]
+  {\bf S}_{\dot a} D_\tau {\bf S}_{\dot a}
+ R \, {\bf S}_{\dot a} [A_9, {\bf S}_{\dot a}]
- 2 i R \,
{\bf X}^i \sigma^i_{a \dot a} \{ {\bf S}_a , {\bf S}_{\dot a} \}
\hskip0.1cm \Big).
\label{lag1}
\eee
We take a convention that all the fields are Hermitian matrices.
This heterotic M(atrix) theory is invariant under eight dynamical and eight
kinematical supersymmetries. In terms of the conjugate momenta ${\bf \Pi}_i$
and $E_9$ to ${\bf X}^i$ and $A_9$ respectively,
the eight dynamical supersymmetry charges  are given by
\be
{\bf Q}_a = {\sqrt R} \, {\rm Tr} \Big[ \Big( {\bf \Pi}_i \sigma^i_{a \dot a}
{\bf S}_{\dot a} + E_9 {\bf S}_a \Big)
+ {1 \over 2} \,
\Big( \hskip0.3cm {\bf X}^i \sigma^{ij}_{ab} [{\bf S}_b, {\bf X}^j]
+ [A_9, {\bf X}^i] \sigma^i_{a \dot a} {\bf S}_{\dot a} \hskip0.3cm \Big)
\hskip0.2cm \Big].
\label{supercharge}
\ee
It is easy to see that the ${\bf Q}_a$ supercharges map
symmetric and adjoint representation fields of $SO(2N)$ back into
symmetric and adjoint representation fields.
For the other eight broken supersymmetries, it is not even
possible to write down a gauge invariant expression of the `would-be'
supercharges ${\bf Q}_{\dot a}$. This is an expected result: while the
covering space M(atrix) theory allows $SU(2N)$ gauge invariant
${\bf 8}_c$ supercharges, orbifold projection reduces the gauge group to
$SO(2N)$ with the above field content. In terms of the new field content,
 the ${\bf 8}_c$ supercharges become no longer gauge invariant.

While the above heterotic M(atrix) theory is perfectly supersymmetric,
the orbifold projection apparently does not lead to equal numbers of bosonic
and spinor degrees of freedom. Counting on-shell degrees of freedom,
there are
$2N(2N-1)/2 + 8 \cdot 2N(2N+1)/2 - 2N(2N -1)/2 = 16 N^2 + 8N$ bosons and
$4 \cdot 2N(2N-1)/2 + 4 \cdot 2N(2N+1)/2 = 16 N^2$ spinors.
As alluded previously, this originates from the excess projection of the
spinor coordinates along the diagonal entries in the ${\bf 8}_s$.
On the other hand, the deficit fermionic degrees of freedom can be matched
by adding eight (plus eight mirror) spinors in the fundamental representation
of the $SO(2N)$ gauge group for each orbifold fixed points, which we call the
`twisted sector'. The twisted sector spinors are neutral under the surviving
supersymmetries of the orbifold projection. As such, each of the untwisted and
the twisted sectors are manifestly supersymmetric.

While the deficit of degrees of freedom seems unavoidable consequence of
the orbifold projection, it is gratifying that the M(atrix) theory tells
us indirectly what the field contents of the twisted sector consists of.
The deficit spinor degrees of freedom are attributed to the ${\bf 8}_s$
representation. As such, the fermions in the twisted sector have quantum
numbers $({\bf 8}, 2N)$  under $SO(8)$ global and $SO(2N)$ gauge symmetries.
Given these quantum numbers, interaction of the twisted sector fermions to
the untwisted sector is uniquely determined~\footnote{A similar twisted sector
fermion interaction was introduced previously in
Ref.~\cite{kachrusilverstein}. Our derivation offers a detailed account
for the M(atrix) theory origin and dynamics governing them.}:
\be
L_{\rm twisted} = \sum_{a=1}^8 \sum_{A,B = 1}^{2N}
{\overline \chi}^A_{1a} ( D_\tau + \gamma_9 R A_9 )_{AB} \chi^B_{1a}
+
{\overline \chi}^A_{2a}
\big(D_\tau + \gamma_9 R (A_9 - i \pi R_9 \sigma_2)
\big)_{AB} \chi_{2a}^B.
\label{twisted}
\ee
In this expression, the two sets of fermions $\chi_1$ and $\chi_2$
originate from each of the two fixed points on
 ${\bf S}_1/{\bf Z}_2$ orbifold.
We also have introduced a notion of `chirality' operator
$\gamma_9$ so that $\gamma_9 \chi_a^A \equiv - \chi_a^A$,
which amounts to a choice of relative
sign between the $A_0$ and $A_9$ field couplings in Eq.(~\ref{twisted}).
This `chirality' operation is motivated by the observation that the twisted
sector fermions originate from to the deficit ${\bf 8}_s$ degrees of freedom
in the untwisted sector.

The $\gamma_9$ `chirality' prescription given
 as above is not arbitrary but
follows from consistency conditions.
First, the twisted sector interactions should be invariant under the residual
supersymmetry. With $\gamma_9 = -1$, the supersymmetry variations of $A_0$
and $A_9$ cancel out each other. Second, as will be shown below, with
$\gamma_9 = -1$ `chirality', the twisted sector fermions become the sixteen
worldsheet fermions whose Kac-Moody currents generate the spacetime gauge
symmetry in the T-dual heterotic string.

The complete heterotic M(atrix) theory Lagrangian is a sum of Eq.(\ref{lag1})
and Eq.(\ref{twisted}). Upon T-duality, the heterotic M(atrix) theory is
related to the D-string in type I string theory. The latter was
interpreted as the heterotic string theory itself~\cite{polchinskiwitten}.
As such, it would be interesting to see how the worldsheet theory of the
heterotic string may be recovered from the heterotic M(atrix) theory.
The T-duality  turns the heterotic M(atrix) theory into an $(1+1)$-dimensional
$SO(2N)$ gauge theory on a dual orbifold:
\bee
L_{(1+1)} =  \oint {d\sigma \over 2 \pi} \,
\Big( \!\!\!\!\!\! &&
\!\!\!\!\!\! {\rm Tr} \,
\Big( \, - {1 \over 4 R} F_{\alpha \beta}^2 +
{1 \over 2R} (D_\alpha {\bf X}^i)^2
+  {R \over 4} [{\bf X}^i, {\bf X}^j]^2
\nonumber \\
&- &  {\bf S}_a D \hskip-0.24cm / {\bf S}_a +
{\bf S}_{\dot a} D \hskip-0.24cm / {\bf S}_{\dot a}
- 2 i R {\bf X}^i \sigma^i_{a {\dot a}} \{ {\bf S}_a, {\bf S}_{\dot a} \}
\, \Big) \nonumber \\
& + & \sum_{a=1}^8 \sum_{A=1}^{2N}
{\overline \chi}^A_{1a} {{\cal D} \hskip-0.24cm /}_{AB} \chi^B_{1a}
+
{\overline \chi}^A_{2a} {{\cal D} \hskip-0.24cm /}_{Ab} \chi^B_{2a}
\hskip0.5cm \Big),
\label{hetworldsheet}
\label{2daction}
\eee
where $i,j = 1, \cdots, 8$ and all the fields depend on
$(\tau, \sigma = R X_9)$. The covariant derivatives are
defined as $D_\alpha {\bf X}^i \equiv (D_\tau{\bf X}^i, D_\sigma {\bf X}^i)$,
$D_\sigma \equiv \partial_\sigma -i [A_9, ]$,
$D \hskip-0.24cm / {\bf S}_a \equiv
(D_\tau + \Gamma^9 D_9) {\bf S}_a$ etc. and ${\cal D}_{AB}
\equiv (D_\tau - D_9)_{AB}$ respectively. We have also made a finite
shift $A_9 \rightarrow A_9 + i \pi R_9 \sigma_2$ by making the boundary
conditions of $\chi_2$ fermions opposite to those of $\chi_1$.

Consider the simplest case $N=1/2$, viz. a single D-parton attached to the
orientifold. Let us take the `light-cone gauge' $A_0 = A_9$.
There are no components to fields of adjoint representations,
while fields of symmetric representations
are $(1 \times 1)$ fields, ${\bf X}^i = X^i(\sigma, \tau) \in {\bf 8}_v$ and
${\bf S}_{\dot a} = S_{\dot a}(\sigma, \tau) = S_{\dot a} (\tau - \sigma)
\in {\bf 8}_c$.
They are the eight bosonic and eight spinor spacetime coordinates of the
Type I D-string in the Green-Schwarz formalism.
In the `light-cone gauge', noting the $\gamma_9$ sign choice of
the twisted sector fermions, $\chi^A_{1,2} = \chi^A_{1,2} (\sigma, \tau)
= \chi^A_{1,2} (\tau + \sigma)$, we find precisely the worldsheet
structure of the heterotic string in which the spacetime supersymmetry is
realized in the light-cone Green-Schwarz formulation and the spacetime
gauge symmetry are generated by worldsheet Kac-Moody current algebra of
the 16 + 16 twisted sector fermions from each fixed points. Noting that
they have opposite boundary conditions each other, the resulting gauge
symmetry is expected to be $E_8 \times E_8$ rather than $SO(32)$.

The necessity of the twisted sector consisting of 16 spinors may be
understood from another point of view as well. First, without the
twisted sector fermions, it is straightforward to see that there arise
non-trivial renormalization of moduli space metric and
potential to the heterotic M(atrix) theory \sl at one loop \rm.
Non-vanishing {\sl static} potential represents M(atrix) theory
manifestation of non-vanishing cosmological constant, viz.
uncancelled dilaton tadpole amplitude. The presence of one-loop induced
cosmological
constant can be probed by scattering a 0-brane parton off the orientifold.
The orbifold projection arranges an `image' 0-brane at the
opposite side of the orientifold. Therefore, integrating out massive modes,
we should expect to reproduce the result of two 0-brane scattering
proportional to $v^4/r^7$. In particular, for a consistent description
of heterotic M(atrix) theory, a 0-brane parton {\sl at rest} should be a
stable configuration, viz. {\sl static} potential between the 0-brane
parton and its image should be absent. The relevant
calculation has been considered in Ref.~\cite{ferretti}. For a 0-brane
`head-on' collision to one of the orientifold, the potential
\footnote{We have checked that the same result follows also for the
`light-cone gauge' $A_0 = A_9 = (i / 2) \sigma^2 v t$.} in the
Born-Oppenheimer approximation is calculated by integrating out massive
modes. In the background gauge
$A_0 = 0$ and $A_9 = (i / 2) \sigma^2 \, v t$,
massive excitations consist of 16 bosonic modes from ${\bf X}^i$ and
8 complex fermionic modes from ${\bf S}_{\dot a}$.
The 0-brane potential due to orbifold fixed point is calculated from 
one-loop Feynman integral (vacuum tadpole) over these massive modes. 
Using proper-time regularization of the one-loop integral, we find:
\bee
{\cal A}_{\rm orientifold} &=& \int_{-\infty}^{+\infty} d t \,\, {\cal V}_{\rm orientifold} (v, v t)
= {\det}^{-8} ( - \partial_\tau^2 + v^2 \tau^2) \,
{\det}^{+4} \left( \begin{array}{cc} \partial_\tau & +  v \tau \\
+ v \tau & \partial_\tau \end{array} \right),
\nonumber \\
{\cal V}_{\rm orientifold} (v, r)
&=& v \int_0^\infty {ds \over \sqrt{\pi s}}
e^{- s r^2} {4 - 2 \cos vs \over \sin vs} =
[-4 r + {v^2 \over 2 r^3} + \cdots].
\eee
The result displays a linear, {\sl velocity-independent}, one-loop
vacuum energy, which arises due to the fact that boson and spinor coordinates
are under different $SO(2N)$ representations, hence, mismatch of boson--fermion
degrees of freedom. The linear potential is not compatible with the
expected stability for a 0-brane parton at rest at a distance away
from the orientifold. In the closed string channel, this is
an exchange of dilaton between the fixed point and the 0-brane parton,
viz. dilaton tadpole amplitude induced by non-vanishing local cosmological
constant. However, the cosmological constant is cancelled {\sl locally}
by the twisted sector fermions \footnote{
It is in fact sufficient to cancel the cosmological constant and dilaton
tadpole amplitude only {\sl globally}. In that case, however, it is not
possible to identify field content of the twisted sector from the heterotic
M(atrix) theory. Moreover, the only stable configuration of 0-brane
partons is only when they are all located at the orbifold fixed point.}
 at each location along $A_9 = (i / 2) \sigma^2 r$.
It is important to recall that the twisted sector fermions are in the
fundamental representation of the M(atrix) gauge group $SO(2N)$~\footnote{
We disagree the result of Ref.~\cite{ferretti}, in which the twisted
sector fermions were apparently taken to be in symmetric
representation, not in fundamental representation. This error has caused
the author of Ref.~\cite{ferretti} to conclude incorrectly that
the total number of twisted sector fermions to be sixteen instead of
thirty-two.}
and the elementary but crucial fact that the distance between probe
0-brane and its mirror is the {\sl twice}
the distance between probe 0-brane and orientifold.
The 0-brane potential due to twisted sector is generated by integrating 
out massive fermionic modes from 16 complex, twisted sector fermions. 
Using proper-time regularization, we find:
\bee
{\cal A}_{\rm twist} &=&
\int_{-\infty}^{+\infty} dt \, {\cal V}_{\rm twist} (v, vt)
= {\det}^{+8} \left( \begin{array}{cc} \partial_\tau &  {v \tau \over2}
\\ {v \tau \over 2} & \partial_\tau \end{array} \right),
\nonumber \\
{\cal V}_{\rm twist} (v, r) &=& - 
{v \over 2} \int_0^\infty {ds \over \sqrt{\pi s}}
e^{- s{r^2\over 4}} {4 \cos {vs \over 2} \over \sin {vs \over 2}}
= [+ 4r + {4 v^2 \over 3 r^3} + \cdots ].
\eee
From the above local analysis, we clearly see that each orientifolds carry
8 (plus 8 `image') negative units of D8-brane RR charge.
It shows that the twisted sector consists of eight fermion
representing D8-branes and their `mirror' images has to be located at the
orientifold location in order to cancel {\sl local} cosmological constant.

A few remarks are in order.
Once the twisted sector spectra is determined as above, moduli space of
twisted sector can be further explored by deformation and displacement
of each of the D8-branes.
In addition,
while the above result is essentially a requirement of dilaton tadpole
cancellation, we also expect to reach the same conclusion from the
anomaly cancellation requirement of $(1+1)$-dimensional worldsheet gauge
theory~Eq.(\ref{hetworldsheet})
 of the T-dual Type I D-string that was identified
as a heterotic string~\cite{polchinskiwitten}. Noting that ${\bf S}_a$ and
${\bf S}_{\dot a}$ transform as left-handed adjoint and right-handed
symmetric representations, we find that the $SO(2N)$ gauge anomaly
$8 C_2 ({\tt adjoint}) - 8 C_2 ({\tt symmetric}) = 8 ((N+2) - (N-2))
C_2 ({\tt fundamental})$ due to untwisted sector fermions
is cancelled precisely by thirty-two twisted sector fermions transforming
as left-handed fundamental representations.
Finally, even though the {\sl local} cosmological
constant is cancelled completely
by the twisted sector fermions at the orbifold fixed point, there
still remains a nontrivial one-loop effect: moduli
space metric of the 0-brane partons is renormalized nontrivially both by the
orientifold and the twisted sector fermions to
$g_{99} = (1 + 11 R/3 r^3)$.
\subsection{Large--N Twisted and Untwisted Two-Branes}
By construction, the heterotic M(atrix) theory Eqs.(\ref{lag1},
\ref{twisted}) is a gauge theory of area-preserving diffeomorphism relevant
to open two-branes. It is therefore of interest whether macroscopic matrix
two-brane configurations can be constructed in the large N limit.

The
simplest open two-brane of direct relevance to the strongly coupled heterotic
string is a cylindrical two-brane whose ends are attached to the two
nine-branes. We have found such a two-brane configuration as follows.
Let the open two-brane is extended along 1-9 direction in which the
1-direction is infinitely extended. It is useful to recall the block matrix
structures of the  $SO(2N)$ adjoint and symmetric representations. Noting
that the block diagonal matrices $Y, Z$ are $(N \times N)$ Hermitian matrices,
a matrix open two-brane configuration is easily constructed as:
\be
{\bf X}_1 =  {R_1 \over \sqrt 2} \left( \begin{array}{cc}
P & 0 \\ 0 & P^{\rm T} \end{array} \right)
, \hskip0.5cm
A_9 = { R_9 \over \sqrt 2} \left( \begin{array}{cc}
Q & 0 \\ 0 & - Q^{\rm T} \end{array} \right),
\label{openmembrane}
\ee
where $P, Q$ are $(N \times N)$ ($N \rightarrow \infty)$ Hermitian matrices
satisfying a commutator $[Q, P] = i $. The configuration yields
\be
{1 \over R_1 R_9} [A_9, {\bf X}_1] = {1 \over 2} [Q, P] \oplus [-Q^{\rm T}, P^{\rm T}]
= {i \over 2} {\bf I}_{2N \times 2N}.
\label{openconfig}
\ee
Due to the reflection symmetry of $A_9$ configuration, the range of
eigenvalues for $A_9, {\bf X}_1$ is $[0, 2 \pi R_1] \otimes [0, \pi R_9]$.
By permuting the 9-th and the 11-th (longitudinal) directions, the two ends
of cylindrical two-brane become precisely the heterotic string propagating
at each nine-branes.

In the background of the above open two-brane, it is also possible for
the twisted sector fermions to show a nontrivial, macroscopic configuration.
Since $A_0 = 0$ for the macroscopic matrix two-brane, the equation of motion
for $\chi^A_a$ coupled to the two-brane is given by:
\be
\Big[ \left( \begin{array}{cc} \partial_\tau & 0 \\ 0 & \partial_\tau
\end{array} \right)  + i
\left( \begin{array}{cc}
Q & 0 \\ 0 & - Q^T \end{array} \right) \Big] \left(
\begin{array}{c} \chi^{(1)}_a \\  \chi^{(2)}_a \end{array} \right) = 0
,
\label{twistfermionequation}
\ee
where we have used the chirality property $\gamma_9 \chi^A_a \equiv - \chi^A_a$
and decomposed the spinors explicitly into 8N partons $\chi^{(1)}_a$
and their mirror image partons $\chi^{(2)}_a$.
Since the heterotic string as the boundary of open two-brane is extended along
1-direction, it is convenient to represent $P =  \sigma$, a c-number
variable on a range $[0, 2 \pi]$, and $Q = i \partial_\sigma$.
In this representation, by combining the eight original and the eight
mirror fermions, a nontrivial configuration is found straightforwardly:
\be
\chi^I = E^I \, \exp [ i (\tau + \sigma)]
\label{twistfermionsolution}
\ee
for a non-vanishing, normalized lattice vector $E^I \in SO(16)$.
The configuration represents twisted sector fermions propagating chirally at
each end of the open two-branes, viz. heterotic string~\footnote{
The solution is valid, however,  strictly
in the large N limit. At finite but large N, the Dirac equation is a set of
2N-coupled equations, which can be solved iteratively. The large N limit
then reduces to the above solution.}.
Similar construction is possible for a disk or a cylindrical
two-brane attached to either of
the two nine-branes. Even though topologically not stable, at least for
the ones with a macroscopic size, classical description should be valid.

In addition to the twisted open two-branes, the heterotic M(atrix) theory
should possess macroscopic, matrix two-branes of closed topology. In fact,
it is straightforward to recognize that the heterotic M(atrix) theory contains
a subsector with sixteen supersymmetry charges. Recall that the orbifold
projection in the heterotic M(atrix) theory has reduced the supersymmetry
into  half that of the covering space M(atrix) theory. The
`would-be' ${\bf 8}_c$ supercharges were no longer $SO(2N)$ gauge-invariant.

On the other hand, away from the orbifold fixed points, strings connect among
partons but not to their images. In terms of the block matrix structures given
in Eqs.(4,5) of section 2, only the diagonal block $(N \times N)$ matrices
describe the parton interactions. These diagonal block matrices are, however,
{\sl unconstrained} $(N \times N)$ Hermitian matrices of $SU(N) \in
SO(2N)$ for both adjoint and symmetric representations.
Once all the matrix fields are truncated to the diagonal block matrices,
it is then easy to recognize that the heterotic M(atrix) theory Eq.(\ref{lag1})
reduces to a $SU(N)$ M(atrix) theory (and its mirror) in which all the fields
are adjoint representations of $SU(N)$.
This subsector should preserve sixteen supersymmetries. The ${\bf 8}_c$
supercharges are $SU(N)$ gauge invariant even though non-invariant under
$SO(2N)$, hence, were projected out in the latter.
Since this subsector is equivalent to the $SU(N)$ M(atrix) theory, closed,
oriented two-branes should be present as well. For definiteness, consider
a toroidal two-brane extended along (1-2) directions. The matrix configuration
is given by:
\be
{\bf X}_1 =  R_1
\left( \begin{array}{cc} Q & 0 \\ 0 & Q^{\rm T} \end{array} \right);
\hskip0.5cm
{\bf X}_2 =  R_2
\left( \begin{array}{cc} P & 0 \\ 0 & P^{\rm T} \end{array} \right).
\label{closedmembrane}
\ee
in the notation given above.
Hence, the two-brane wrapping number is found:
\bee
[{\bf X}_1, {\bf X}_2]
&=& i \sigma_3 \otimes {\bf I}_{N \times N}, \nonumber \\
{\cal Z}_{2 \times 2} &=& {1 \over N} {\rm Tr}_{N \times N}
\Big(i [X_1, X_2] \Big) = \left( \begin{array}{cc} 1 & 0 \\ 0 & -1
\end{array} \right).
\eee
The relative negative sign between the two-brane and its image correctly
counts for the orientation reversal of the two-brane or, equivalently, of the
antisymmetric tensor field $A_{MNP} \rightarrow - A_{MNP}$ under the
orbifold condition.

\subsection{Large N Limit}
We finally discuss the large N limit of the heterotic M(atrix) theory.
In this limit, we expect that the M(atrix) theory becomes a $(2+1)$
dimensional field theory coupled to a $(1+1)$-dimensional boundary field
theory.
Since the heterotic M(atrix) theory is constructed from an involution of
$SU(2N)$ M(atrix) theory, we first recapitulate the large N limit recipe
of the latter and then consistently impose the involution condition of
$SO(2N)$ M(atrix) theory.

In this section, extending results of Ref.~\cite{dewit,dewit2,zachosetal}
we explain correspondence rules between the matrix fields
and the continuum fields in the large N limit for open two-branes.
For definiteness, we will consider cylindrical involution in what follows.

In the large N limit of M(atrix) theory, Hermitian $SU(N)$ matrices are
expanded as ${\bf X}^I = \sum_{\bf m} X^I_{\bf m} J_{\bf m}$, where
$J_{\bf m}$ is a complete set of basis for the $(N \times N)$
matrices introduced in Eq.(\ref{matrixbasis}) of section 3.
Therefore, it is sufficient to study large N limit of the basis matrices
$J^\dagger_{\bf m} = J_{- \bf m}$.
In section 3.2.1, we have shown that $J_{\bf m}$'s satisfy the
trigonometric Lie algebra Eq.(\ref{matrixcommutator}). In the large N
limit, the trigonometric Lie algebra is reduced to
\be
[{\bf T}_{\bf m}, {\bf T}_{\bf n}]
= ({\bf m} \times {\bf n}) {\bf T}_{\bf m + n}
\label{largenmatrixalgebra}
\ee
for ${\bf T}_{\bf m} \equiv (N / 2 \pi i) J_{\bf m}$. This algebra is
most conveniently described in terms of two non-commuting variables
${\hat p}, {\hat q}$ satisfying commutation relation
$
[{\hat q}, \, {\hat p}] = {2 \pi i \over N}
$
so that
\be
U = e^{ i {\hat p}}, \hskip0.3cm V = e^{i {\hat q}};
\hskip0.5cm J_{\bf m} = e^{ i m {\hat p} + i n {\hat q}}.
\label{pqrealization}
\ee
Since the non-commutativity is suppressed in the large N limit,
the matrices $J_{\bf m}$ approach ordinary functions that depend on the
classical phase--space variables ${\bf p} = (p,q)$:
\be
J_{\bf m} \rightarrow Y_{\bf m} ({\bf p})
\equiv \exp (i {\bf m} \cdot  {\bf p}).
\ee
In the phase--space, the Poisson bracket algebra of $Y_{\bf m}$'s
\be
\{ Y_{\bf m}, Y_{\bf n} \}_{\rm PB}
= ({\bf m} \times {\bf n}) Y_{\bf m + n}
\ee
becomes isomorphic to the large N commutator algebra
Eq.(\ref{largenmatrixalgebra}).
Therefore, by introducing doubly periodic functions
$X^I({\bf p}) \equiv \sum_{\bf m} X^I_{\bf m} Y_{\bf m}({\bf p})$,
in the large N limit, the matrix commutator algebra can be replaced by the
Poisson bracket algebra:
\be
[{\bf X}_I,{\bf X}_J] \rightarrow \frac{2\pi i}{N}
\{X_I({\bf p}),X_J({\bf p}) \}_{\rm PB}.
\label{commutatoralgebra}
\ee

Next, we replace trace over matrices into a phase space integral. To
do so, it is convenient to recast the trace operation as an inner product:
\be
\tr {\bf X}_I^\dagger {\bf X}_J \equiv \left< {\bf X}_I | {\bf X}_J \right>.
\ee
While we consider Hermitian matrices exclusively, we find it more convenient
to retain the Hermitian conjugation operation and keep track of the
inner product structure. Thus,
\be
\tr J^\dagger_{\bf m} J_{\bf k} = \tr J_{- \bf m} J_{\bf k}
= N \, \delta_{\bf m, k}.
\ee
On the other hand, for the classical basis functions on the classical phase
space, we have
\be
\oint_{-\pi}^{\pi} dp \oint_{-\pi}^{\pi} dq \,\,
Y^*_{\bf m}({\bf p}) Y_{\bf k}({\bf p}) = (2\pi)^2 \delta_{\bf m, k}.
\ee
Thus we are led to the following identification
\bee
\tr {\bf X}_I^\dagger {\bf X}_J^{} & \rightarrow & \frac{N}{(2\pi)^2}
\oint_{-\pi}^{\pi} dp
                           \oint_{-\pi}^{\pi} dq X^*_I({\bf p}) X_J({\bf p}).
\label{trace}
\eee

Finally, consider the $N$-dimensional fundamental representations coupled to
matrices, relevant to the twisted sector.
Fundamental representations $\chi^A$ of $SO(N)$ are most conveniently
expanded
\be
\chi^I = \sum_{n} \chi^I_n {\bf e}_n
\ee
in terms of the basis vectors:
\begin{equation}
{\bf e}_1 = \pmatrix{1 \cr 0 \cr 0 \cr \vdots \cr 0}, \quad
{\bf e}_2 = \pmatrix{0 \cr 1 \cr 0 \cr \vdots \cr 0}, \quad \ldots \quad ,
{\bf e}_N = \pmatrix{0 \cr 0 \cr 0 \cr \vdots \cr 1}.
\end{equation}
The action of $SO(N)$ matrix basis $U, V$ in section 3, Eq.(29)
to the basis vectors is given by
\begin{eqnarray}
U \, {\bf e}_n &=& \omega^n {\bf e}_{n}, \nonumber \\
V \, {\bf e}_n &=& {\bf e}_{n+1}
\end{eqnarray}
with an identification ${\bf e}_{N+n}={\bf e}_{n}$. This algebra,
with an inner product $<{\bf e}_k|{\bf e}_l>\equiv {\bf e}^T_k
\cdot {\bf e}_l=\delta_{kl}$,
can be represented by periodic functions and operators acting on them
defined on a periodic interval $p \in [-\pi,\pi]$:
\begin{eqnarray}
e^{i\hat{q}} \, \exp(imp) &=& \omega^m \exp(imp), \nonumber \\
e^{i\hat{p}} \, \exp(imp) &=& \exp (i(m+1)p), \nonumber \\
\int_{-\pi}^{\pi} dp \, (e^{ikp})^* e^{imp} &=& 2\pi \delta_{km},
\end{eqnarray}
where $\hat{q}\equiv - \frac{2\pi i }{N} \partial_p$ and $\hat{p}
\equiv p$. Thus, identifying $U = e^{\frac{2\pi}{N}\partial_p}$,
$V= e^{ip}$, and ${\bf e}_n= e^{inp}$, we obtain the large N limit
expresssion of twisted sector interactions:
\begin{equation}
\chi^I {\bf X}_{IJ} \chi^J \rightarrow \oint_{-\pi}^{\pi} dp \,
\chi^*(p) X (p, -\frac{\partial}{\partial p}) \chi (p)
\label{twist}
\end{equation}
at each boundaries $q = 0, \pi$.

The basis of heterotic M(atrix) theory Eq.(28) of section 3 restricts
the continuum fields to $ q \in [ 0, \pi]$ with two fixed points
at $q = 0, \pi$. As such, the continuum fields should be
supplemented by boundary conditions. In the large N limit, the orbifold
conditions section 2, Eq.(1) are replaced by reflection conditions:
\bee
X_\perp (p,q) &=& - X_\perp (p, -q), \nonumber \\
X_{\parallel} (p, q) &=& + X_{\parallel} (p, -q), \nonumber \\
S_a (p,q) &=& - S_a (p, -q), \nonumber \\
S_{\dot a} (p,q) &=& + S_{\dot a} (p, -q).
\eee
Therefore, the appropriate boundary conditions are
\bee
X_\perp (p, q=0, \pi) &=& 0, \nonumber \\
\partial_q X_{\parallel} (p, q=0, \pi) &=& 0, \nonumber \\
S_a (p, q=0, \pi) &=& 0, \nonumber \\
\partial_q S_{\dot a} (p, q=0, \pi) &=& 0.
\label{boundarycondition}
\eee
The boundary condition agrees  with the ones derived from the M-theory
\cite{beckerbecker}.

With the above correspondence rules Eqs.(\ref{commutatoralgebra},
\ref{trace}, \ref{twist}) and the boundary conditions
Eq.(\ref{boundarycondition}), it is straightforward to find continuum
expression of the heterotic M(atrix) theory Lagrangian.

\section{Discussions}
M(atrix) theory as a viable non-perturbative definition of M-theory has
passed various consistency tests so far~\cite{banksetal2, susskind, taylor,
aharonyberkooz, berkoozdouglas}. In particular, it has been shown
that the M(atrix) theory supports
two-brane~\cite{banksetal} and longitudinal five-brane~\cite{banksetal2},
which are shown to be BPS states made out of infinitely many 0-branes.

In this paper, we have initiated investigation of M(atrix) theory on an
orbifold. We have shown that different choices of M(atrix) theory gauge
group give rise to different topologies of classical two-branes. This is
not surprising after all: the M(atrix) theory is nothing but a gauge theory
of area-preserving diffeomorphism transformation. Interestingly, while
Chan-Paton analysis constrains the possible choices of M(atrix) theory
gauge groups, we have found that it is through the area-preserving
diffeomorphism analysis that we uncover more refined information of the
M(atrix) theory. We have found that, among $SO(2N)$, $SO(2N+1)$ and $USp(2N)$
gauge groups allowed by Chan-Paton factor analysis for M(atrix) theory, only
$SO(2N)$ or $SO(2N+1)$ M(atrix) theory is capable of describing nontrivial
twisted two-branes as BPS excitations.

As the simplest yet non-trivial example, we have considered the heterotic
M(atrix) theory. This is a M(atrix) theory defined on an orbifold
${\bf S}_1/{\bf Z}_2 $. We have shown that Bose-Fermi degrees of freedom
matching, hence, cancellation of localized cosmological constants require
an introduction of twisted sector consisting of 16 fermions (plus
mirrors) transforming
in the fundamental
representation of $SO(2N)$ or $SO(2N+1)$ M(atrix) gauge group.
They are nothing but the
8 D8-branes (plus mirrors) located at each orbifold fixed points.
We have constructed a classical BPS configuration of open two-brane stretched
between the two fixed points. The two ends are nothing but the heterotic
string moving freely in the ten-dimensional spacetime. In addition, we also
found a closed two-brane hovering around the full eleven-dimensional spacetime.
By studying the large N limit, we have shown that the correct boundary
conditions to the open two-brane are obtained for both bosonic and spinor
coordinate fields.

In the forthcoming papers~\cite{kimrey},
we will report consistency checks of the
heterotic M(atrix) theory and orbifold M(atrix) theories in higher-dimensional
orbifolds, especially, $({\bf S}_1)^5/{\bf Z}_2$ and $({\bf S}_1)^9/{\bf Z}_2$
related by string dualities to compactified string theories.

\vskip0.5cm
We thank M. Dine, D. Kabat, E. Silverstein
and L. Susskind for discussions and I. Bars and M.R.
Douglas for correspondences.


\end{document}